\documentclass[bm,aps,
amsfonts,amssymb,preprint,nofootinbib]{revtex4}

\usepackage{graphicx}
\usepackage{natbib}
\usepackage{amsmath}

\usepackage{color}

\font\mybb=msbm10 at 12pt

\def\bb#1{\hbox{\mybb#1}}

\def\PP {\bb{P}}

\newcommand\beqa{\begin{eqnarray}}
\newcommand\eeqa{\end{eqnarray}}
\newcommand\n{\nonumber\\}

\begin{document}

{~}

\title{Unitary matrix 
with a Penner-like 
potential\\also yields $N_f=2$}
\vspace{2cm}
\author{Shun'ya Mizoguchi\footnote[1]{E-mail:mizoguch@post.kek.jp},
Hajime Otsuka\footnote[2]{E-mail:hotsuka@post.kek.jp} 
and Hitomi Tashiro\footnote[3]{E-mail:tashiro@post.kek.jp}}

\vspace{1cm}

\affiliation{\footnotemark[1]\footnotemark[2]Theory Center, 
Institute of Particle and Nuclear Studies,
KEK\\Tsukuba, Ibaraki, 305-0801, Japan 
}

\affiliation{\footnotemark[1]\footnotemark[3]SOKENDAI (The Graduate University for Advanced Studies)\\
Tsukuba, Ibaraki, 305-0801, Japan 
}

\begin{abstract} 
It has been known for some time that a hermitian matrix model 
with a Penner-like potential yields as its large-$N$ free energy 
the prepotential of ${\cal N}=2$ $N_f=2$ $SU(2)$ SUSY gauge theory. 
We give a rigorous proof that a {\em unitary} matrix model with 
the identical potential also yields the same prepotential, although 
the parameter identifications are slightly different. This result 
has been anticipated by Itoyama et.~al.

\end{abstract}

\preprint{KEK-TH-2156}
\date{September 19, 2019}

\maketitle

\newpage
\section{Introduction}
When the Dotsenko-Fateev integral representation of correlation 
functions of two-dimensional CFT was derived 35 years ago \cite{DotsenkoFateev}, and also when  
the double-scaling limit of the one-matrix model was discovered 
30 years ago \cite{BrezinKazakov,DouglasShenker,GrossMigdal}, 
no one would have imagined that they would be related to any 
four-dimensional theory. However, in the early 2000's 
the relationship between 4d SUSY gauge theories and matrix models 
was revealed \cite{DijkgraafVafa1,DijkgraafVafa2,DijkgraafVafa3},
later the AGT relation between 4d SUSY gauge theories and 
2d CFT was found \cite{AGT}, and finally it turned out that 
their correspondence was elegantly formulated in terms of 
matrix models \cite{DijkgraafVafa2009,MironovMorozov2009,
Itoyama:2009sc,EguchiMaruyoshi1,Schiappa:2009cc,
Mironov:2009ib,MMS2010}.

Once having recognized the connection between 4d ${\cal N}=2$ 
SUSY gauge theories and matrix models, the origin of the mysterious 
appearance of the Painlev\'e equations \cite{BrezinKazakov,DouglasShenker,GrossMigdal} 
in the double-scaling limit
of the latter is now seen to be natural; 
the total space of the Seiberg-Witten curve 
(including the base ``$u$-plane''(= an affine patch of $\PP^1$) of 
the elliptic fibration) of an 
${\cal N}=2$ $SU(2)$ SUSY gauge theory (as well as 
an $E$-string)  
can be identified as 
a rational elliptic surface\cite{MV2,Ganor1,Ganor2,KMV,GMS,MNVW,YY}, 
and it is these 
particular algebraic varieties that the Painlev\'e equations 
were shown to be associated with in 2001 \cite{Sakai}. 
The latter was a geometric manifestation of the idea of 
constructing discrete Painlev\'e equations as translations of 
affine Weyl groups \cite{NoumiYamada}.  
Recently, there has been an interest in the Painlev\'e equations 
in SUSY gauge theories in terms of irregular conformal blocks 
and double-scaled matrix models.  Recent works in this direction include  
\cite{ItoyamaOotaYonezawa2010,GIL12,GIL13,ILT,BGT1603,BGT1704,
nag1611,BLMST,GG,LNR1806,
AJJRT1607,ItoyamaOotaYano1805,ItoyamaOotaYano1812}.

In \cite{ItoyamaOotaYano1805}, Itoyama, Oota and Yano 
claimed, among other things,  
that a {\em unitary} one-matrix model with a logarithmic potential 
term yields the instanton partition function 
\cite{Nekrasov,NakajimaYoshioka} of the 
${\cal N}=2$ $N_f=2$ $SU(2)$ SUSY gauge theory.  
Prior to this, it had been explicitly confirmed by Eguchi and Maruyoshi 
\cite{EguchiMaruyoshi2}
that a {\em hermitian} one-matrix model with the same potential 
reproduces the instanton partition function of the above 
same gauge theory as its large-$N$ free energy. 
This is quite puzzling because, even if they have the same 
form of the potential, they are a priori {\em different} matrix 
models with {\em different} Boltzmann weights defined by 
different integration contours of the eigenvalues. That is, 
in hermitian matrix models, the contour is taken to be 
the real axis, whereas in unitary matrix models the 
eigenvalue integration is performed 
along the unit circle around the origin. 
Also, one cannot expect to be able to change the contour 
as one does in a residue computation of a holomorphic 
integral since the integrand function is not holomorphic,  
and even has a logarithmic singularity on the real axis 
in the hermitian case.

In this paper, we examine, by an explicit calculation, 
whether Itoyama et.al.'s unitary matrix model with 
a logarithmic potential can yield the prepotential of 
the $N_f=2$ theory. Our strategy is as follows: we 
first map Itoyama et.al.'s unitary matrix model to 
an equivalent hermitian matrix model giving the 
identical partition function by using the 
unitary/hermitian duality in matrix models 
\cite{MizoguchiUnitaryHermitian,BowickMorozovShevitz}. 
We then compute its two-cut large-$N$ free energy 
following the standard techniques and derive 
its matrix model curve. We examine whether we can 
find the $u$-parameter appropriately so that 
the $u$ derivative of the matrix model differential 
($y(z)dz$ in the text) 
becomes proportional to the holomorphic differential. 
If we can find one, then the differential is identified 
(up to a constant of proportionality) as the Seiberg-Witten 
differential whose Seiberg-Witten curve is the 
genus-one Riemann surface associated with the 
holomorphic differential above. The ``special geometry relation'' 
\cite{DijkgraafVafa3,Cachazo:2003yc,FujiMizoguchi,
MizoguchiUnitaryHermitian} 
then automatically ensures that the matrix-model
free energy coincides with the prepotential of 
the gauge theory (up to the term linear in the Coulomb 
modulus $a$).  

Surprisingly, we will see that the unitary 
matrix model of Itoyama et. al. yields, through these 
procedures, precisely {\em the same} Seiberg-Witten 
curve and Seiberg-Witten differential as those obtained 
in the hermitian matrix model 
having the potential of the same form 
analyzed in \cite{EguchiMaruyoshi2}! 
Thus this implies that the two different - hermitian and unitary - 
matrix models with the same logarithmic potential 
computes, as their large-$N$ free energy, 
the instanton partition function of an identical 
gauge theory, the ${\cal N}=2$ $N_f=2$ $SU(2)$ theory.

This paper is organized as follows. In section 2, we revisit 
the Penner-like hermitian matrix model studied in \cite{EguchiMaruyoshi2},
where we solve it by using the standard conventional 
technique for solving hermitian matrix models. We reproduce 
the results obtained in \cite{EguchiMaruyoshi2}, such as
the conditions for the positions of the end points of the cuts, 
the Seiberg-Witten curve and the $N_f=2$ prepotential as 
its large-$N$ free energy. We then turn to the unitary matrix model 
with the same potential in section 3. We use the unitary/hermitian 
matrix model duality to convert the unitary matrix model to 
the equivalent hermitian matrix model. Then we solve it similarly 
to find that it also describes $N_f=2$. 
The conclusions are summarized in Section 4.

\section{Hermitian matrix model with a Penner-like potential revisited}
Our convention for the matrix model partition function is 
\beqa
Z&=&\frac1{\mbox{Vol}(U(N))}
\int d\Phi \exp\left(
-\frac N\mu \rm{tr}W(\Phi)
\right),
\eeqa
where $\mu=g_s N$ is the t'Hooft coupling.
The potential is 
\beqa
W(z)&=&-\mu_3\log z +\frac\Lambda 2\left(
z+\frac 1z
\right),
\label{EguchiMaruyoshiW}
\eeqa
where the overall sign is flipped compared to the 
definition in \cite{EguchiMaruyoshi2}. Following the standard technique, 
the resolvent $\omega(z)$ for a 2-cut solution is given by 
\beqa
\omega(z)&=&\frac{\sqrt{\prod_{k=1}^4(z-a_k)}}{4\pi i \mu}
\sum_{j=1,2}\oint_{A_j}dw\frac{W(w)}{(z-w)\sqrt{\prod_{k=1}^4(w-a_k)}},
\label{general2cut}
\eeqa 
where $A_j$ $(j=1,2)$ are the cuts on which the eigenvalues are distributed. 
The definition for the resolvent is
\beqa
\omega_(z)&\equiv&\frac1N\left<
{\rm tr}\frac1{z-\Phi}
\right>.
\eeqa
The positions of the endpoints of the cuts are constrained, 
as usual, by the asymptotic behavior of the resolvent 
\beqa
\omega(z)&\sim&\frac 1z+O\left(
\frac1{z^2}
\right).
\eeqa 
Let
\beqa
\prod_{k=1}^4(z-a_k)&\equiv&z^4+b_1 z^3 + b_2 z^2 + b_3 z + b_4,  
\eeqa
then
by Laurrnt-expanding (\ref{general2cut}) around $z=\infty$, 
we obtain the constraints
\beqa
\frac1{2\mu}\left(
-\frac{\mu_3}{\sqrt{b_4}}+\frac{\Lambda b_3}{4\sqrt{b_4}^3}
\right)&=&0,\\
\frac1{2\mu}\cdot\frac{b_1}2\left(
-\frac{\mu_3}{\sqrt{b_4}}+\frac{\Lambda b_3}{4\sqrt{b_4}^3}
\right)
+\frac\Lambda{4\mu}-\frac{\Lambda}{4\mu\sqrt{b_4}}&=&0,\\
-\frac{\mu_3}{2\mu}+\frac1{2\mu}\left(
\frac{b_2}2-\frac{b_1^2}8
\right)
\left(
-\frac{\mu_3}{\sqrt{b_4}}+\frac{\Lambda b_3}{4\sqrt{b_4}^3}
\right)
-\frac{\Lambda}{2\sqrt{b_4}}\cdot\frac{b_1}{4\mu}&=&1,
\eeqa
which reproduce the relations found in \cite{EguchiMaruyoshi2}:
\beqa
b_3=\frac{4\mu_3}\Lambda,~~~b_4=1,~~~
b_1=-\frac{8\mu+4\mu_3}\Lambda.
\eeqa

The large-$N$ expansion:
\beqa
Z&=&\exp\left(\frac{N^2}{\mu^2} F\right),
\\
F&=&F_0+\frac{\mu^2}{N^2} F_1+\cdots.
\eeqa
$F_0$ is the large-$N$ free energy given by the well-known formula:
\beqa
\frac1{\mu^2}F_0&=&-\frac1\mu\int_{-\infty}^\infty d \lambda
\rho(\lambda)W(\lambda)
+
\int_{-\infty}^\infty d \lambda
\int_{-\infty}^\infty d \lambda'
\rho(\lambda)\rho(\lambda')
\log|\lambda-\lambda'|,
\eeqa
where $\rho(\lambda)$ is the eigenvalue density.

In the present case
\beqa
\omega(z)&\equiv&\frac1{2\mu}(W'(z)- y(z)),\\
y(z)&=&\frac\Lambda 2\sqrt{1+\frac{b_1}z+\frac{b_2}{z^2}+\frac{b_3}{z^3}+\frac1{z^4}}.
\label{hermitian_y(z)}
\eeqa
Let 
\beqa
\mu_1&\equiv&-\frac1{4\pi i}\oint_{A_1}y(z),
\eeqa
then $F_0$ satisfies the ``special geometry relation'', 
meaning that
\beqa
\frac{\partial F_0}{\partial \mu_1}&\propto&
\oint_B y(z),
\eeqa
where $B$ is the other homology cycle. Therefore, 
if there exists some parameter variable $u$ such that 
$\frac{\partial y(z)}{\partial u} dz$ is proportional to the 
holomorphic differential, $y(z)dz$ is essentially the Seiberg-Witten 
differential of the curve. In the present case, we can take $b_2$ as $u$, 
then (up to a term linear in $\mu_1$) 
$F_0$ is automatically the prepotential of the special K\"ahler 
geometry.

\section{Unitary matrix model with a Penner-like potential}
The partition function of a unitary matrix model is 
defined as usual by
\beqa
Z&=&\frac1{\mbox{Vol}(U(N))}
\int dU \exp\left(
-\frac N\mu {\rm tr}W_U(U)
\right)\\
&=&\int \prod_{i=1}^N\left(
d\theta_i e^{-\frac N\mu W_U(e^{i\theta_i})}
\right)
\prod_{j<k}^N \sin^2 \frac{\theta_j-\theta_k}2.
\eeqa

We take the same potential (\ref{EguchiMaruyoshiW})
as the potential for the unitary matrix model here. 
Then this is a Penner-like generalization of the Gross-Witten-Wadia 
model \cite{GrossWitten,Wadia1,Wadia2}.
For convenience, we absorb the factor $\frac\Lambda 2$ 
by an overall rescaling and shift $\theta_i$ by $\frac\pi 2$.
The potential we consider is  
\beqa
W_U(U)&=i^{-1}(U-U^{-1})+\nu \log(-U).
\eeqa
If $\nu=0$, it reduces to the Gross-Witten-Wadia model though 
the potential is $\sin$ instead of $\cos$ due to the shift. 

It has been shown that a unitary matrix model with 
a potential $W_U(U)$ is equivalent \cite{MizoguchiUnitaryHermitian,BowickMorozovShevitz} 
to a hermitian matrix 
model with a potential 
\beqa
W(\Phi)&=&W_\Phi(\Phi)+\mu\log(1+\Phi^2),
\eeqa
where
\beqa
W_\Phi(\Phi)&\equiv&W_U\left(
\frac{i-\Phi}{i+\Phi}
\right).
\eeqa
In the present case, we have
\beqa
W(\Phi)&=&\frac{4\Phi}{1+\Phi^2}+\mu_+\log(\Phi -i)
+\mu_-\log(\Phi+i),
\eeqa
where $\mu_\pm\equiv\mu\pm\nu$.

\beqa
W'(x)&=&\frac{4(1-x^2)}{(1+x^2)^2}+\frac{\mu_+}{x-i}+\frac{\mu_-}{x+i}.
\eeqa

\beqa
\omega(z)&=&\frac{\sqrt{\prod_{k=1}^4(z-a_k)}}{4\pi i \mu}
\left(
\oint_{x=\infty}-\oint_{x=z}-\oint_{x=i}-\oint_{x=-i}
\right)dx
\frac{\frac{4(1-x^2)}{(1+x^2)^2}+\frac{\mu_+}{x-i}+\frac{\mu_-}{x+i}}
{(z-x)\sqrt{\prod_{k=1}^4(x-a_k)}}.
\label{omega_integral_form}
\eeqa

\beqa
\mbox{The 1st term}&=&0,\n
\mbox{The 2nd term}&=&\frac1{2\mu}W'(z),\n
\mbox{The 3rd term}&=&
\frac{\sqrt{\prod_{k=1}^4(z-a_k)}}{2\mu(i-z)\sqrt{\prod_{k=1}^4(i-a_k)}}
\left(
\frac2{i-z}+\sum_{k=1}^4\frac1{i-a_k}+\mu_+
\right),\n
\mbox{The 4th term}&=&
\frac{\sqrt{\prod_{k=1}^4(z-a_k)}}{2\mu(-i-z)\sqrt{\prod_{k=1}^4(-i-a_k)}}
\left(
\frac2{-i-z}+\sum_{k=1}^4\frac1{-i-a_k}+\mu_-
\right).
\eeqa

Again, the positions of the end points of the branch cuts 
are not arbitrary, but are constrained in order for 
the resolvent $\omega(z)$ to have a correct asymptotic 
behavior at infinity. Expanding (\ref{omega_integral_form})
around $z=\infty$, we find the conditions
\beqa
\prod_{k=1}^4(i-a_k)&=&\prod_{k=1}^4(-i-a_k),
\label{constraint1}
\\
\sum_{k=1}^4\frac1{i-a_k}+\mu_+&=&
-\left(\sum_{k=1}^4\frac1{-i-a_k}+\mu_-\right)~=~-2i.
\label{constraint2}
\eeqa
Thus, defining
\beqa
\omega(z)&\equiv&\frac1{2\mu}(W'(z)-y(z)),
\eeqa
we obtain
\beqa
y(z)&=&\frac{8\sqrt{\prod_{k=1}^4(z-a_k)}}{A(z^2+1)^2},
\label{y(z)}
\eeqa
where
\beqa
A&\equiv&\sqrt{\prod_{k=1}^4(i-a_k)}~=~\sqrt{\prod_{k=1}^4(-i-a_k)}.
\eeqa
Note that the explicit dependence of $\mu_\pm$ 
disappears in (\ref{y(z)}), but they affect $y(z)$ only 
through $a_k$'s by the relation (\ref{constraint2}).  

To see that $y(z)dz$ is a Seiberg-Witten differential for some $u$, 
let us go back to the unitary-matrix complex coordinate 
by the replacements
\beqa
z&=&\frac{i(1-w)}{1+w}
\eeqa
and 
\beqa
a_k&=&\frac{i(1-T_k)}{1+T_k}~~~(k=1,\cdots,4).
\label{aktoTk}
\eeqa
In terms of $w$ and $T_k$'s, $y(z)dz$ becomes
\beqa
y(z)dz&=&\frac{-i\sqrt{\prod_{k=1}^4(w-T_k)}}{w^2} dw.
\eeqa
If we write
\beqa
\prod_{k=1}^4(w-T_k)&\equiv&w^4+\sigma_1 w^3
+\sigma_2 w^2+\sigma_3 w+\sigma_4, 
\eeqa
then by using (\ref{aktoTk}) in (\ref{constraint1}) and (\ref{constraint2}) 
we find 
\beqa
\sigma_4=1,~~\sigma_3=2i\mu_+,~~\sigma_1=-2i\mu_-.
\eeqa
Thus we finally obtain
\beqa
y(z)dz&=&\frac{\sqrt{{\hat w}^4+2\mu_-{\hat w}^3-\sigma_2 {\hat w}^2
+ 2\mu_+ {\hat w}+1}}{{\hat w}^2} d {\hat w}\label{y(z)dzunitary}\\
&\equiv& \hat y(\hat w)d \hat w \nonumber,
\eeqa
where ${\hat w}\equiv i w$. This is precisely the same expression  
as $y(z)$ (\ref{hermitian_y(z)}) of the hermitian matrix model 
we saw in the previous section, up to an overall constant factor.
Again, we can take $-\sigma_2$ as $u$, then
\beqa
\frac{\partial \hat y(\hat w)d \hat w}{\partial u}&=&\frac{d {\hat w}}
{2\sqrt{{\hat w}^4+2\mu_-{\hat w}^3+u {\hat w}^2
+ 2\mu_+ {\hat w}+1}}\\
&\equiv&\frac{d \hat w}{2v(\hat w)} \nonumber,
\eeqa
which is a holomorphic differential of the $N_f=2$ curve
\beqa
v^2&=&{\hat w}^4+2\mu_-{\hat w}^3+u {\hat w}^2
+ 2\mu_+ {\hat w}+1.
\eeqa
It has been shown \cite{MizoguchiUnitaryHermitian} 
that, even in the presence of the $\log$ potential terms, 
the periods of the differential (\ref{y(z)dzunitary}) satisfy
\beqa
\displaystyle\hskip -48pt \frac{\partial F_0}{\partial \mu_j}
&=&
\frac12 \lim_{\epsilon\rightarrow 0}\left[
\int_{\widetilde{-i+i\epsilon}} ^{i-i\epsilon}
\!\!\!\!\!\!\!\!\!\mbox{\scriptsize ($j$th cut) }
dz y(z)
~-W(i-i\epsilon)-W(\widetilde{-i+i\epsilon})\right],
\label{dF0dmuj}\\&&
\oint_{A_j} dz y(z)\equiv-4 \pi i \mu_j ~~(j=1,2),
\label{yAperiod}
\eeqa
where $~\widetilde{}~$ denotes the point on the second sheet 
specified by the value of $z$.
The relation (\ref{dF0dmuj}) holds in the case where 
the two periods $\mu_1$ and $\mu_2$ are treated as independent
variables. If, instead, $\mu_1$ and $\mu(=\mu_1+\mu_2)$ are treated 
as independent, then 
\beqa
\left.\frac{\partial F_0}{\partial \mu_1}
\right|_{\mbox{\scriptsize $\mu$}}&=&
\left.\frac{\partial F_0}{\partial \mu_1}
\right|_{\mbox{\scriptsize $\mu_2$}}-
\left.\frac{\partial F_0}{\partial \mu_2}
\right|_{\mbox{\scriptsize $\mu_1$}}\nonumber\\
&=& -\frac12 \oint_B dz y(z),
\eeqa
which shows that the unitary-matrix free energy $F_0$ 
is the prepotential of the $N_f=2$ (up to a term linear in $\mu_1$).

\section{Conclusions}

We have shown that a {\em unitary} matrix model with a Penner-like potential 
identical to the one used in Eguchi-Maruyoshi's hermitian matrix model 
also yields the prepotential of ${\cal N}=2$ $N_f=2$ $SU(2)$ SUSY gauge 
theory as its large-$N$ free energy, the fact anticipated by Itoyama et.~al. 

Although the two matrix models describe the same gauge theory, 
the parameter identifications are slightly different. For instance, in 
the hermitian model, the coefficient of the $\log$ term in the potential 
corresponds to one of the mass of the flavor, while in the unitary model, 
the coefficient of $\log$ represents the {\em difference} of the masses of 
the two flavors. (This fact was already noted in \cite{ItoyamaOotaYano1805}.) 
After all, we can say that the two matrix models are 
different models with different Boltzmann weights, but they can describe 
the same gauge theory if the parameter identifications are changed 
in a suitable way.

Itoyama et.~al.'s proposal  \cite{ItoyamaOotaYano1805}
that the Penner-like unitary matrix model also yields $N_f=2$ was 
concluded by considering an irregular limit of the AGT relation,  
in which they used a few assumptions. 
(See Appendix F of \cite{ItoyamaOotaYano1812}.)
Our result implies the validity of the assumptions.

Finally, we would like to emphasize that what we have shown in this paper 
implies more than the fact the unitary and hermitian matrix models 
belong to the same universality class 
\cite{PeriwalShevitz1,PeriwalShevitz2} 
in the critical limit.
Our proof holds true even away from the criticality, though, of course, 
being in the same class will be a necessary condition for their equivalence.

We thank H.~Itoyama and T.~Oota for useful discussions and 
correspondence. We also thank N.~Kan, 
K.~Maruyoshi and K.~Sakai for discussions.
The work of H. O. was supported by Grant-in-Aid for JSPS Fellows No. 19J00664. 

\end{document}